\documentclass[prl,twocolumn,letterpaper]{revtex4}
\usepackage{epsfig}
\usepackage{amsmath,amssymb}
\usepackage{graphicx}

\begin{document}

\title{A Sudden Gravitational Transition}
\author{Robert R. Caldwell$^1$, William Komp$^2$, 
Leonard Parker$^3$, Daniel A.\ T.\ Vanzella$^4$}
\affiliation{
$^1$ Department of Physics and Astronomy, Dartmouth College, 6127 Wilder 
Laboratory, Hanover, NH 03755 USA\\
$^2$ Physics Department, University of Louisville, 102 Natural Sciences,
Louisville, KY 40292 USA\\
$^3$ Physics Department, University of Wisconsin-Milwaukee, P.O. Box 413, 
Milwaukee, Wisconsin 53201 USA\\
$^4$ Instituto de F\'\i sica de S\~ao Carlos, Universidade de S\~ao Paulo (IFSC-USP), 
Av. Trabalhador S\~ao-carlense, 400 Cx. Postal 369 - CEP 13560-970,
S\~ao Carlos, S\~ao Paulo, BRAZIL
}

\date{\today}

\begin{abstract} 

We investigate the properties of a cosmological scenario which undergoes a
gravitational phase transition at late times. In this scenario, the Universe
evolves according to general relativity in the standard, hot Big Bang picture
until a redshift $z \lesssim 1$. Non-perturbative phenomena associated with a 
minimally-coupled scalar field catalyzes a transition, whereby an order
parameter consisting of curvature quantities such as $R^2,\, R_{ab}R^{ab},
\,R_{abcd}R^{abcd}$ acquires a constant expectation value. The ensuing cosmic
acceleration appears driven by a dark-energy component with an equation-of-state
$w < -1$. We evaluate the constraints from type 1a supernovae, the cosmic
microwave background, and other cosmological observations. We find that a range
of models making a sharp transition to cosmic acceleration are consistent with
observations.

\end{abstract}
 
\maketitle


\noindent{\it Introduction.} The observational evidence for a low density,
spatially-flat, accelerating Universe poses a severe challenge to theoretical
physics. Solutions which have been proposed include a cosmological constant, an
ultra-light scalar field, and modifications to Einstein's gravity. An approach
which unifies these different viewpoints can be found in Sakharov's description
of the gravitational physics of the vacuum \cite{Sakharov:1967pk}. According to
his proposal, quantum effects of the particles and fields present in the
Universe give rise to the cosmological constant and gravitation itself. Although
this approach has not solved the problem of the cosmological constant, it has
pointed the way towards a novel model for the dark energy.

Quantum effects of an ultra-light, minimally-coupled scalar field have been
proposed to account for the dark energy phenomena in the vacuum metamorphosis scenario
\cite{Parker:1999td,Parker:1999ac,Parker:2000pr,Parker:2001,Parker:2004}.
Effectively, the Ricci scalar curvature serves as an order parameter, marking a 
gravitational transition when it drops to the value $\chi m^2$.  Here, $m \sim
10^{-33}$~eV is the mass of the scalar field and $\chi$ is a numerical constant
of order unity which is set by the theory. For most of the history of the
Universe, up until $z \sim 1$, $R > \chi m^2$ and the vacuum stress-energy is
negligble. Since the local value of the Ricci scalar exceeds this critical value
in the vicinity of galaxies today, we see no vacuum energy nearby. On the largest
scales, however, the vacuum term comes to dominate the cosmic evolution as $R \to
\chi m^2$. A detailed analysis shows that feedback from the vacuum prevents $R$
from evolving past the critical value, in a sort of gravitational Lenz' law which
maintains $R= \chi m^2$ \cite{Parker:2004}. The ensuing, large-scale cosmic
expansion begins to accelerate, driven by a dark energy component with an
equation-of-state $w < -1$ which approaches $-1$ in the future.  

Despite the sharp change in the character of gravitation, we are still justified
to use basic cosmological tools such as the luminosity distance. Although the
transition changes the relationship between curvature and matter, we still have
a metric theory and a complete description of the evolution of the expansion
scale factor.

The vacuum metamorphosis scenario is distinct from scalar-tensor theories of
gravity \cite{Yasunori} or high-energy physics inspired modifications of the
gravitational action whereby  the Einstein-Hilbert action is replaced by a
function of the Ricci scalar, $-16 \pi G {\cal L}_{g} = R \to f(R)$. (See
Ref.~\cite{Carroll:2003wy} for a recent example.) In our case, the
gravitational action is modified by vacuum polarization effects of the
matter content in the theory. However, the modifications include
contributions from $R,\,R_{ab},\,R_{abcd}$ so there is not a simple way to
re-express the model in terms of an equivalent, non-minimally-coupled scalar
field. (See Refs.~\cite{Teyssandier:1983,Wands:1993uu} for a discussion of
the equivalence between higher-order gravity theories and scalar-tensor
gravity.)   Furthermore, the scenario we consider is different from other
dark energy models based on curved-space quantum field theory effects 
\cite{Sahni:1998at,Onemli:2004mb}, in that the cosmic acceleration arises
from non-linear, non-perturbative phenomena. In fact, the theory we examine
in this article is similar in spirit to the gravitational transition
investigated by Tkachev \cite{Tkachev:1992an}. The distinction here is in
the form of the vacuum stress-energy, which is based on the novel results by
one of us and collaborators
\cite{Parker:1999td,Parker:1999ac,Parker:2000pr,Parker:2001,Parker:2004}.
The one-loop effective action, which is complete for a free massive scalar
field in curved spacetime, can be viewed perturbatively in terms of Feynman
diagrams. Classical gravitons attach to the vacuum loops of the scalar
field. With increasing number of external gravitons, these diagrams require
counterterms that give rise to renormalization of the cosmological constant,
Newton's constant, and several other constants that appear in terms that are
of second order in the Riemann tensor. In the limit of a massless field, the
conformal trace anomaly arises from finite parts of the second order terms
and does not depend on the values of the renormalized constants. The terms
in the perturbative expansion that are of still higher order in the Riemann
tensor have no infinities and appear to be relevant only at Planckian
scales. However, summation of an infinite subset of those terms shows that
they may give rise to a large nonperturbative effect when the Ricci scalar
curvature approaches a specific value proportional to the square of the
scalar particle's mass. Alternatively, the functional integral over
fluctuations of the scalar field in the one-loop effective action can be
performed and the result can be expressed in terms of the exact heat kernel
of the scalar field equation. By studying known exact solutions for the heat
kernel, one can infer a plausible asymptotic form of the heat kernel and
show that it leads to the same type of large nonperturbative effect when the
Ricci scalar curvature approaches a value, $\chi m^2$, proportional to the
square of the particle's mass. This latter approach does not rely on a
subset of Feynman diagrams and is thus fully nonperturbative. Like the
conformal trace anomaly, this effect is independent of the values of
renormalized constants. If the calculation is valid, vacuum metamorphosis
must occur if the universe contains a light scalar field with $\chi$
positive. Since a complete description of the stress-energy tensor source
for Einstein's equations must include all matter fields as well as vacuum
stress-energy, the cosmological phenomena resulting from the vacuum
stress-energy is inevitable given the existence of such a scalar field. 

The original vacuum metamorphosis model is tightly constrained, although not
eliminated, by observations \cite{Parker:2003,Komp:2004}. We find the basic
behavior to be of sufficient interest to justify further investigation. In this
article we speculate that other curvature quantities such as $R_{ab}R^{ab}$ or
$R_{abcd}R^{abcd}$ can serve a similar role as order parameters for a
gravitational transition. In the following we examine the cosmic evolution
resulting from such gravitational transitions. We evaluate the observational
constraints based on type 1a supernovae, the cosmic microwave background,  the
Hubble constant, and large scale structure. We stop short of making a full
perturbation analysis --- the detailed equations require a lengthy investigation
--- and we regard this as a first cut at a family of models extending the vacuum
metamorphosis scenario.

\vspace{0.2cm}

\noindent{\it Gravitational Transition.}
In the vacuum metamorphosis scenario, non-perturbative effects of a light scalar
field lead to a gravitational transition when the Ricci scalar reaches the level
$R = {\bar m}^2$. Here we absorb the dimensionless parameter $\chi$ into the
mass $\bar m$. Before the transition, the cosmic evolution is determined by the 
standard FRW equation
\begin{equation}
3 H^2 = 8 \pi G  \rho_m \qquad a < a_*
\label{FRWbefore}
\end{equation}
where $a_*$ marks the time of the transition, and $\rho_m$ represents all matter
and radiation. There is no need to include the  vacuum energy density, $\rho_v$,
since it is negligible at these early times. After the transition, however, the
cosmic evolution is given by
\begin{equation}
R = 6 (\dot H + 2 H^2) = {\bar m}^2 \qquad a \ge a_* \, .
\label{vcdmafter}
\end{equation}
Notably, the subsequent evolution of matter and radiation after the transition
have no influence on the expansion rate. The vacuum energy does not merely
contribute to the cosmic energy density driving the expansion, as for most dark
energy models. Rather, it changes the form of gravity on cosmological scales and
completely determines the expansion. After solving (\ref{vcdmafter}) for $H(a)$
or $a(t)$, however, we can still use the standard FRW equation
\begin{equation}
3 H^2 = 8 \pi G \left(\rho_{m} + \rho_{v}\right)
\label{FRWafter}
\end{equation}
together with the evolution of the matter density parameter
\begin{equation}
\Omega_m(a)
= \left(\frac{H_*}{H}\right)^2 \left(\frac{a_*}{a}\right)^3 \, ,
\label{omegam}
\end{equation}
to deduce the properties of an equivalent dark energy that has the
equation-of-state, $w = p_v / \rho_v$:
\begin{equation}
w = - \left(1 + \frac{2}{3}{\dot H /H^2}\right)/\left(1 - \Omega_{m}(a) \right).
\label{wdark}
\end{equation}
The inferred equation-of-state is super-negative, with $w < -1$ which approaches
$w \to -1$ in the future. For this model, there is a single free parameter,
$\bar m$. So, after choosing a present day Hubble constant $H_0$, the value of
$\bar m$ determines the matter density:
\begin{equation}
\Omega_m|_0 = \frac{2 \sqrt{2}}{3} \left(\frac{\bar m}{H_0}\right)^2
\left[ \left(\frac{H_0}{\bar m}\right)^2 - \frac{1}{12}\right]^{3/4} \, .
\end{equation}
The Hubble constant at the time of the transition is $H_* = \bar m / \sqrt{3}$,
so 
\begin{equation} 
H(a) = \frac{\bar m}{2 \sqrt{3}} \left[3\left(\frac{a_*}{a}\right)^4 +
1\right]^{1/2}
\end{equation}
gives the analytic solution to the evolution equation (\ref{vcdmafter}). Note
that the total energy density and pressure is always positive: $\rho + p = \bar
m^2 (a_*/a)^4 / 8 \pi G$. There is no ``big rip" in this scenario
\cite{Caldwell:2003vq}, as the late-time evolution resembles a comological
constant-dominated universe.
 

We propose to extend this gravitational transition model, using $R_{ab}R^{ab}$
or $R_{abcd}R^{abcd}$ as the order parameter in place of $R$. In either case,
the evolution before the transition is determined by the standard FRW equation
(\ref{FRWbefore}).  After the transition, from the constancy of the order
parameter we have
\begin{equation}
\alpha\left[ \dot H^2 + N( \dot H H^2 +  H^4)\right] = \bar m^4 \qquad a \ge a_*
\label{constcurv}
\end{equation}
where $\alpha=12$ and $N=2,\,3$ for the Riemann and Ricci tensors, and 
$\alpha=36,\, N=4$ for the Ricci scalar. Linear combinations of these curvature
terms which lead to  cosmic acceleration, $\ddot a > 0$, correspond to  $3/2 < N
\le 9/2$. We can solve these equations numerically to determine the subsequent
evolution.

\begin{figure}[ht]
\includegraphics[scale=0.95]{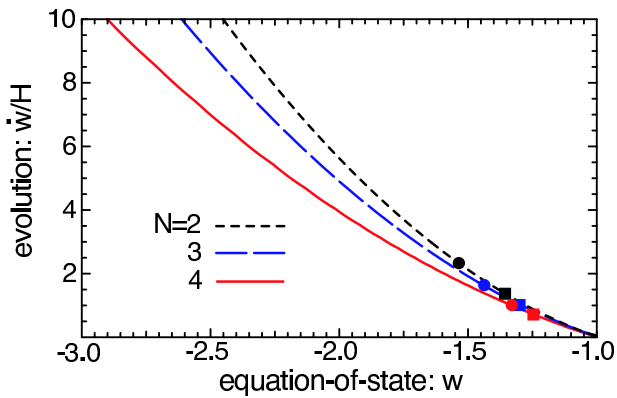}
\caption{The evolution of the gravitational transition model in the $w-\dot w/H$
phase plane is shown.  For a given $N$, the model trajectory follows the curve
indicated with $\Omega_m$ decreasing as the phase variables evolve from the
starting point at $(w,\,\dot w/H)=(-\infty,\infty)$ and head towards $(-1,0)$.
The values corresponding to $\Omega_m=0.3,\,0.4$ are indicated by the small
squares and circles, respectively. In the case $\Omega_m=0.4$ today, $(w,\dot
w/H)=(-1.55,2.4),\,(-1.45,1.6), \,(-1.35,1.0)$,  with the transition occuring
at  $a_*/a_0 = 0.7,\,0.6,\ 0.5$   for $N=2,\,3,\,4$ respectively.}
\label{fig0a}
\end{figure}

\begin{figure}[ht]
\includegraphics[scale=0.95]{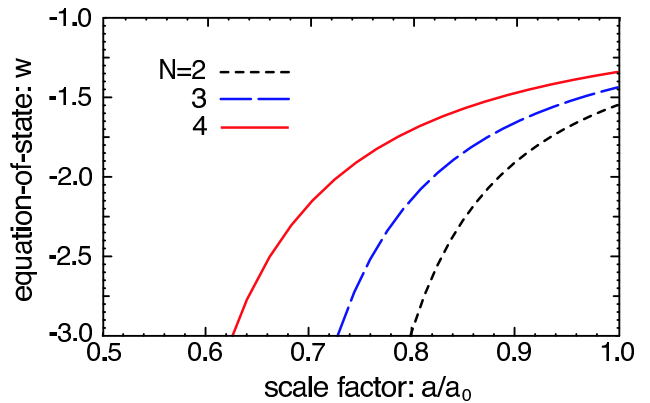}
\caption{The evolution of the equation-of-state $w$ as a function of the scale
factor $a$ is shown. All models have $\Omega_m=0.4$ today. For decreasing $N$,
the effective gravitational repulsion of the dark energy as measured by $w$
increases. The $N=4$ model is consistent with all observations, the $N=3$ is on
the border, and $N=2$ is excluded.}
\label{fig0b}
\end{figure}

\begin{figure}[ht]
\includegraphics[scale=0.95]{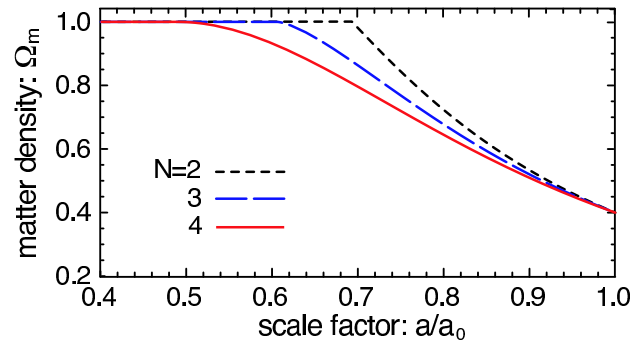}
\caption{The evolution of the density parameter $\Omega_m$ as a function of the
scale factor $a$ is shown. All models have $\Omega_m=0.4$ today. The matter
density drops suddenly at the onset of the transition. The sharpness of the drop
grows with decreasing $N$. The $N=4$ model is consistent with all observations,
the $N=3$ is on the border, and $N=2$ is excluded.}
\label{fig0c}
\end{figure}

The cosmic evolution under various gravitational transition scenarios is shown
in Figures~\ref{fig0a},\ref{fig0b},\ref{fig0c}. The hallmark of these models is
a rapidly-evolving equation-of-state with $w < -1$. We can see that by
decreasing $N$, or increasing the rank of the curvature quantities used to form
the order parameter, the strength of the vacuum energy increases. The total
$\rho+p$ drops faster in time, and the onset of cosmic acceleration is sharper.

Another possibility that comes to mind as an order parameter is the Gauss-Bonnet
invariant, 
\begin{equation}
Q = R^2 - 4 R_{ab}R^{ab} + R_{abcd}R^{abcd}.
\end{equation} 
However, the deceleration parameter $q \equiv -a \ddot a /\dot a^2$ is
proportional to the Gauss-Bonnet invariant, $q = -Q/24 H^4$. This means the
curvature cannot evolve down to some constant $Q$ to signal a gravitational
transition to cosmic acceleration: $Q <0$ in a matter-dominated universe, but
$Q>0$ is required for acceleration.

\vspace{0.2cm}

\noindent{\it Constraints.} 
We now consider the observational constraints on the cosmic evolution resulting
from the family of generalized vacuum metamorphosis models. These constraints
are due to: 1) distance - redshift relationship using type 1a supernovae (SNe)
\cite{Riess:2004nr,Knop:2003iy}; 2) the distance to the cosmic microwave
background (CMB) last scattering surface and the density $\Omega_m h^2$ implied
by the WMAP measurements \cite{Spergel:2003cb}; 3) the Hubble constant based on
the HST key project \cite{Freedman:2000cf}; 4) the mass power spectrum shape
parameter $\Gamma \equiv \Omega_m h$ \cite{Percival:2001hw,Pope:2004cc}.

\begin{itemize}
\item For the SNe we use the 156 supernova, ``gold" data set of Riess {\it et al}
\cite{Riess:2004nr} and the 54 supernova set of Knop {\it et al}
\cite{Knop:2003iy}. We make a simple $\chi^2$ test to determine the $2\sigma$
region allowed for each data set. 
\item For the CMB we exploit the geometric degeneracy of CDM-family models with
identical primordial perturbation spectra, matter content at last scattering, and
comoving distance to the surface of last scattering 
\cite{Zaldarriaga:1997ch,Bond:1997wr,Efstathiou:1998xx}. This means that there is
a family of dark energy models with different equation-of-state histories and
dark energy abundances but essentially identical CMB spectra \cite{Huey:1998se}.
Hence, our best-fitting models are those which are degenerate with the
best-fitting $\Lambda$CDM models obtained by WMAP  \cite{Spergel:2003cb}. We have
taken the $2\sigma$ $\Lambda$CDM models based on a 5-parameter fit
($\Omega_\Lambda,\, \Omega_c h^2,\, \Omega_b h^2,\, h,\, n_s$ obtained from a
combination of earlier work \cite{Caldwell:2003hz}, new calculations using
CMBfast \cite{Seljak:1996is}, and CMBfit \cite{Sandvik:2003ii}). 
We also use the Big Bang Nucelosynthesis prior $\Omega_b h^2 = 0.02 \pm
0.002\,(95\%)$ \cite{Burles:2000zk}. Ultimately, these points
in the $\Lambda$CDM $\Omega_m-H_0$ plane
are mapped to points in the VCDM $\Omega_m-H_0$ plane by fixing the
quantities
$\Omega_m h^2$ and the luminosity distance to the last scattering surface.
This procedure  overlooks
differences in the large-angle anisotropy pattern, which would require a full
treatment of the cosmological perturbations after the transition to model
accurately. However, the $\chi^2$ fit between the theoretical model and
experiment is dominated by the small-angle anisotropy pattern so these
differences should be small. (A preliminary treatment of the large-angle
CMB anisotropy suggests the VCDM constraint region shifts to higher
$\Omega_m$ by $\lesssim 10\%$ \cite{Komp:2004}.)
\item For the Hubble constant, we require that the value of $H_0$ 
falls within $2\sigma$ of HST's result $72 \pm
8$~km/s/Mpc ($1\sigma$) \cite{Freedman:2000cf}.  
\item For the shape parameter, the chief obstacle in applying the 2dF and SDSS results
is the difference in the rate of growth of linear density perturbations between
our models and a $\Lambda$CDM cosmology. We expect linear perturbation growth to
cease at the transition, although we postpone a detailed treatment for a future
investigation. The current $1\sigma$ bound on $\Gamma$ has a $\sim 20\%$
uncertainty which should be comparable to the differences resulting from the
perturbation growth.  For safe margin, we require that our best-fit model has a
shape parameter which falls within $3\sigma$ of the bound $\Gamma = 0.2 \pm 0.03
~(1\sigma)$ \cite{Percival:2001hw,Pope:2004cc}.
\end{itemize}

\begin{figure}[t]
\includegraphics[scale=0.6]{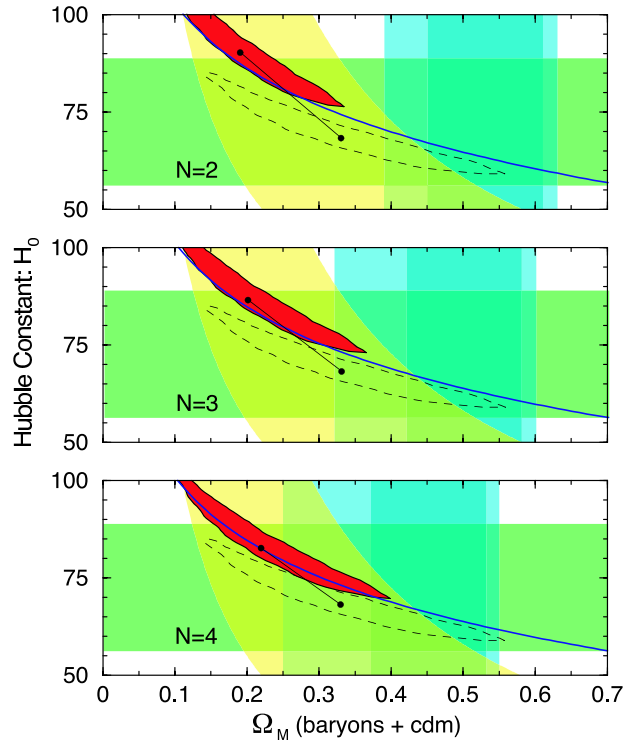}
\caption{The observational constraints on the $\Omega_m - H_0$ parameter space
for the $N=2,\,3,\,4$ gravitational transition models are shown. The red-filled
contour is the WMAP CMB $2\sigma$ region, obtained from the  $\Lambda$CDM
$2\sigma$ region, given by the black dashed contour, by using the geometric
degeneracy. The line connecting the black dots shows the mapping of the
best-fit $\Lambda$CDM model to the gravitational transition model. The SN
$2\sigma$ regions are shown by the wide, pale blue (Knop et al 2004) and
narrow, pale grey (Riess et al 2004) bands. The shape parameter  $\Gamma$
$3\sigma$ region is shown by the yellow swath. The Hubble constant $2\sigma$
region is the horizontal green band. For reference, gravitational transition
models with age $13.5$~Gyrs lie along the thin blue line. There is significant
overlap amongst all model constraints for the cases $N=4$ for the $R^2$
transition. Sharper transitions, corresponding to $N<4$, are in conflict with
some or all of the observations.}
\label{fig1}
\end{figure}  
 
The results are shown in Figure~\ref{fig1}. We see that the gravitational
transition models require a lower matter density than $\Lambda$CDM to satisfy
the CMB constraints, and a higher matter density to satisfy the SN constraints.
The $\chi^2$ fit to the CMB data for the best-fit model is identical to that for
the best-fit $\Lambda$CDM model, due to the geometric degeneracy. The fit to the
SN data is marginally better than $\Lambda$CDM ($\chi^2=175$ for $156$ SNe
\cite{Riess:2004nr};  $\chi^2=59$ for $54$ SNe \cite{Knop:2003iy}). For $N=4$,
the original VCDM model, there are viable models near $\Omega_m = 0.4,\, H_0=70$
whereby the transition redshift is $z_* \sim 1$ \cite{Komp:2004}. However, as
$N$ decreases, the tension between CMB and SN data builds. Hence, sharpening the
gravitational transition relative to VCDM conflicts with observation. As $N$
approaches $9/2$ it turns out that the gap between the SN and CMB parameter
ranges decreases. But in this limit, for fixed matter density today, the
transition occurs at earlier and earlier times, at $z_* \gg 1$. Since linear
perturbation growth is expected to slow or cease at the transition, such an
early transition can be expected to wreck the basic structure formation
scenario. 

There are some straightforward ways to ease the conflict between these
gravitational transition models and observations: 1) allow for a small amount of
spatial curvature; 2) include the energy of the scalar field itself.

\begin{figure}[t]
\includegraphics[scale=0.6]{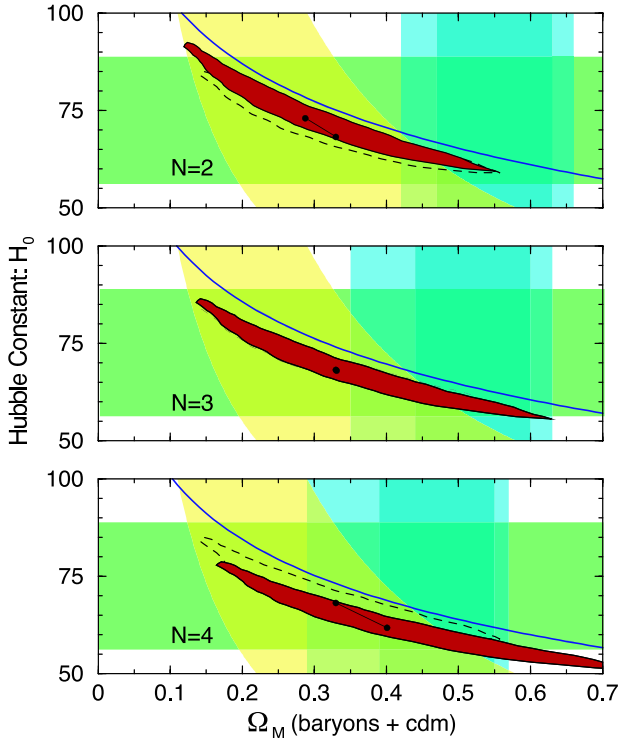}
\caption{Same as Fig.~\ref{fig1} but with positive spatial curvature. In all
cases $\Omega_k=-0.05$ and $\Omega_m=1.05$ at the time of the transition. The
agreement with CMB and SN constraints improves significantly, particularly for
the N=4 case as the CMB best-fit point falls inside the $2\sigma$ contour for
all other constraints.}
\label{fig1k}
\end{figure}

Adding negative spatial curvature improves the fit to the SN data while lowering
the matter density. However, the viable CMB region shifts to an even lower range
of $\Omega_m$, and so the two major constraints remain in conflict. Adding
positive spatial curvature, thereby closing the Universe, brings the $2\sigma$ SN
and CMB regions into closer agreement. As illustrated in Fig.~\ref{fig1k}, the
addition of a small amount of positive spatial curvature, $\Omega_k = -0.05$ at
the transition, leads to concordance. The quality of the improvement is best for
$N=4$, whereby $\Omega_m\approx 0.4$ and $H_0 \approx 60$~km/s/Mpc lies within the
$2\sigma$ contour for all the constraints. Yet, we view the addition of spatial
curvature as extraneous. It is unrelated to the mechanism of vacuum metamorphosis
or gravitational transition, and there is no direct observational evidence which
requires it. Hence, we do not pursue spatial curvature any further.

Including the energy of the scalar field itself can also bring the SN and CMB
constraint regions into better agreement. Here it is necessary to explain that
the vacuum stress-energy which gives rise to the gravitational transition
consists of the gravitational corrections to the scalar field stress-energy in
the state in which the vacuum expectation value of the scalar field is zero. A
non-zero vacuum expectation value would contribute the same stress-energy as a
free classical scalar field of mass $m$.

\begin{figure}[b]
\includegraphics[scale=1.0]{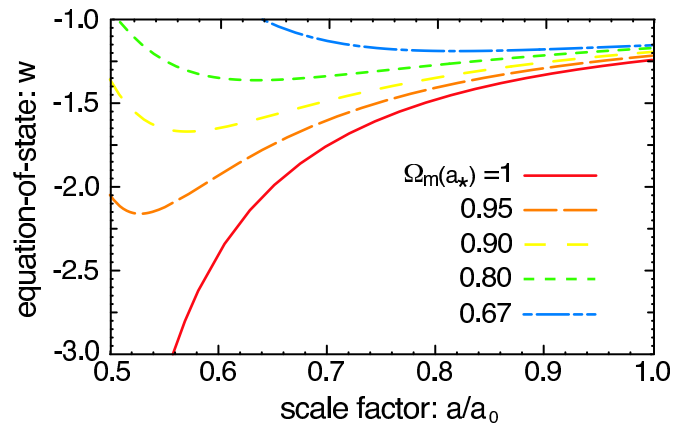}
\caption{The evolution of the equation-of-state $w$ as a function of the scale
factor $a$ is shown for the $N=4$ model including the classical scalar field.
In all cases, $w$ starts at $-1$ at the transition, drops below, and then
ultimately evolves back to $-1$ in the asymptotic future. For $\Omega_m=0.3$
today, only the top curve, with $\Omega_m=0.67$ at the transition,  produces a
viable cosmology satisfying all observational constraints.}
\label{fig0d}
\end{figure}  

This classical scalar field introduces new degrees of freedom into the model.
The mass of the scalar field, with potential $V = \frac{1}{2} m^2\phi^2$, is
related by $m^2 = \bar m^2/\chi$ to the parameter $\bar m$ that determines the
cosmological evolution after the transition. Using the value $\chi=6$ suggested
by earlier versions of the VCDM model \cite{Parker:1999td}, then $m^2 = \bar m^2
/ 6$. The evolution of such a light field will be strongly damped by the cosmic
expansion up to the time of the transition, so that to first approximation it
will look like a cosmological constant.  In order to contribute a fraction $1 -
\Omega_m(a_*)$ of the total cosmic energy density at the transition time, the
field amplitude must be $\phi = \sqrt{3 (1-\Omega_m(a_*))/2 \pi} M_{P}$ where
$M_{P}$ is the Planck mass. This is a potential sticking point for the model,
much as for scalar field quintessence, since such a high field amplitude should
be susceptible to quantum gravitational effects which, for example, induce
couplings to all other matter fields. However, the absence of non-gravitational
interactions for $\phi$, together with its very small mass $m$, appear to
suppress such couplings. Also, one may question whether it is necessary or
economical to introduce a second form of dark energy. This reduces to the
question of the vacuum expectation value of the field, which must be determined
by the initial conditions or perhaps post-inflationary physics. Whether or not
the vacuum expectation value is large, for the quantum scalar field that we have
been considering, the quantum effects that lead to vacuum metamorphosis are
inevitable.

Including an effective cosmological constant before the transition, the
evolution equations for the Hubble expansion rate or the scale factor are the
same as (\ref{constcurv}), but with different initial conditions at the
transition. For the $R^2$ case with $N=4$, the analytic solution is
\begin{eqnarray} 
H(a) &=& \frac{\bar m}{2 \sqrt{3}} 
\left[\frac{3 \Omega_m(a_*)}{4 - 3\Omega_m(a_*)}
\left(\frac{a_*}{a}\right)^4 + 1\right]^{1/2}, \\
\Omega_m(a) &=& \Omega_m(a_*) 
\left(\frac{H_*}{H}\right)^2 \left(\frac{a_*}{a}\right)^3 \, .
\end{eqnarray}
The equation-of-state is still obtained from (\ref{wdark}). But whereas $w$ is
undefined at the transition in the absence of the classical scalar field, now
$w(a_*)=-1$ as illustrated in Figure~\ref{fig0d}. Also, note that
$\Omega_m(a_*) \ge \Omega_m(a_0)$. In the $N=4$ model with $\Omega_m(a_*)=2/3$
and $\Omega_m(a_0)=0.3$, the transition  occurs at $a_*/a_0=0.64$ and the
equation-of-state is $w=-1.15$ today. As $\Omega_m(a_*)$ is lowered, the onset
of cosmic acceleration is earlier but the transition is later, for fixed
$\Omega_m(a_0)$. In the limit that $\Omega_m(a_*) = \Omega_m(a_0)$ the
transition occurs at $z_*=0$ and the model is equivalent to $\Lambda$CDM.

\begin{figure}[t]
\includegraphics[scale=0.6]{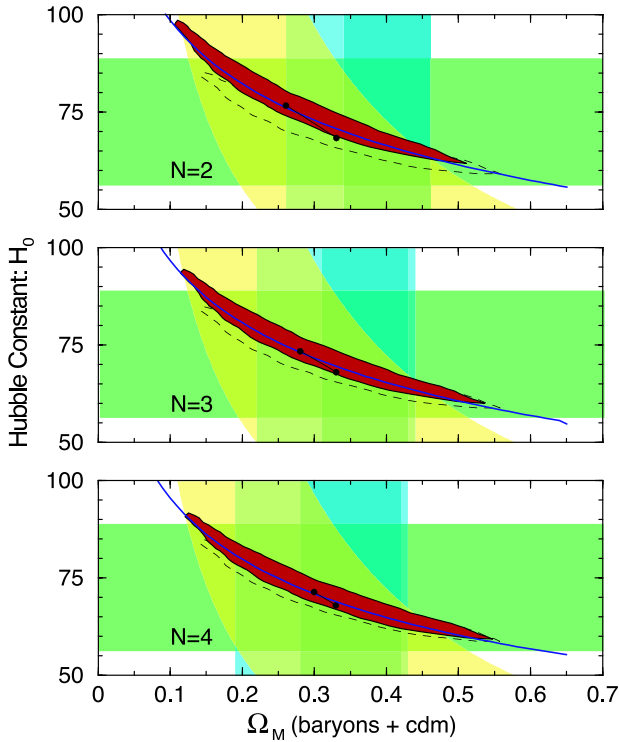}
\caption{Same as Fig.~\ref{fig1}, but the classical mode of the scalar field
contributes an effective cosmological constant before the gravitational
transition. At the transition, $\Omega_m(a_*) = 2/3,\, \Omega_\Lambda(a_*)=1/3$.
The result is that the transition to acceleration is softened, and the
tension between the CMB and SN constraints is relaxed. Models with $N=2-4$ are
viable.} \label{fig2}
\end{figure}

We have evaluated observational constraints for the $R^2,\, R_{ab} R^{ab}$ and
$R_{abcd}R^{abcd}$ models with a cosmological constant. As more $\Lambda$  is
added before the transition, the gap between the CMB and SN constraints reduces.
In Figure~\ref{fig2} we show the constraints resulting for the case
$\Omega_{m}(a_*) = 2/3$. The quality of the $\chi^2$ fit to the CMB and SN data
are approximately the same as for the cases without the addition of the scalar
field contribution. In all cases shown there are viable models near $\Omega_m =
0.3$ and $H_0 = 70$, providing good motivation for further investigation of
these models.

\vspace{0.2cm}
 
\noindent{\it Cosmological Perturbations.} 
The vacuum metamorphosis scenario and its generalizations have dramatic
implications for gravitation.  The transition occurs when the order parameter,
$R^2$ in the $N=4$ case, reaches a critical value. On large scales, when the
cosmological curvature drops to $R=\bar m^2$ the transition takes place. On
sub-horizon scales, on the scales of voids, this critical value is reached a bit
earlier than on average; on the scale of clusters, this value is reached a bit
later.  Since $R = 8 \pi G(\rho+ 3p)$ in Einstein's gravitation, a high density
or pressure will keep $R$ above the critical value. Inside a cluster and on the
scale of galaxies, where the mean density is well above the cosmic background,
the transition never takes place.

What happens to gravitation after the transition? The field equations become
higher-order, so that a static potential obeys a fourth order equation,
$(-A\nabla^2\nabla^2 + \nabla^2) \Phi = 4 \pi B G\rho$. $\sqrt{A}$ is a length,
below which the potential is exponentially suppressed. $B$ modifies the strength
of gravity, and grows tiny upon the transition. This would look like a disaster,
except that the transition takes place only on large scales, where this
Newtonian analysis is invalid.

To treat the scenario in cosmological perturbation theory, we first observe that
the gravitational transition takes place on a spacelike hypersurface of constant
curvature. And the duration of the transition is effectively instantaneous,
based on a numerical modeling of the vacuum effects \cite{Parker:2004}. Taking
into account small cosmological perturbations, this surface will be a surface of
constant $R + \delta R$ to linear order. At times after the transition, the
feedback of the vacuum effects on the gravitational field equations forces $R$
to a constant. It is reasonable to surmise that fluctuations $\delta R$ are
forced to vanish. To deal with the evolution of fluctuations across the
transition we resort to junction conditions, ensuring energy and momentum flow
is continuous. Following Ref.~\cite{Deruelle:1995kd}, we can choose a gauge and
then match the perturbation variables from pre- to post-transition. In general,
we can expect that the pure growing mode which dominates the evolution before
the transition will give way to a linear combination of growing and decaying
modes afterwards. The evolution of the perturbations after the transition,
however, presents a challenge.

If the transition forces fluctuations of the scalar curvature to vanish, then we
find the constraint
\begin{eqnarray}
\delta R &=& 8 \pi G(\delta\rho + 3 \delta p)  \cr
&=&\frac{1}{a^2} \left( h'' + 3 {\cal H} h' - 4 k^2\eta\right) = 0
\end{eqnarray}
which should be valid on the range of scales for which the mean curvature has
frozen at $R=\bar m^2$. Similar constraints arise for the $N=2,\,3$ cases. Here
we work in the synchronous gauge with metric perturbation variables
$h$ and $\eta$, following the notation of
Ref.~\cite{Ma:1995ey}, ${\cal H}=a'/a$, and the prime indicates a derivative
with respect to conformal time. Since density perturbations in the
non-relativistic matter respond to gravitational fluctuations according to
$\delta_m' = -h'/2$, we have
\begin{equation}
\delta_m'' + 3 {\cal H}\delta_m' + 8 k^2\eta = 0 .
\end{equation}
Compared to the standard case, for which $\delta_m'' + 2 {\cal H}\delta_m' + 4
k^2\eta=0$, we expect stronger Hubble damping but also a stronger source. We are
tempted to use one of the perturbed Einstein's equations to replace  $\eta$ with
a fluid variable such as $\delta \rho$ for the density perturbations. However,
we must not forget to include all contributions to the fluid  perturbations. We
know that the vacuum effects must contribute  an equivalent energy density
$\delta\rho_v$ and pressure $\delta p_v$ such  that $\delta\rho_v + \delta\rho_m
+ 3 \delta p_v = 0$. But that's all we know without making a detailed
calculation. (A brief glance at the stress-energy tensor for the vacuum
metamorphosis model \cite{Parker:2004} should convince the reader that this is
not so straightforward.) Yet, the evolution equations for the cosmic expansion
look similarly difficult and a simple result, transition to constant curvature,
is obtained. There may be a simple resolution to the perturbation evolution as
well.

\vspace{0.2cm}

\begin{figure}[t]
\includegraphics[scale=0.5]{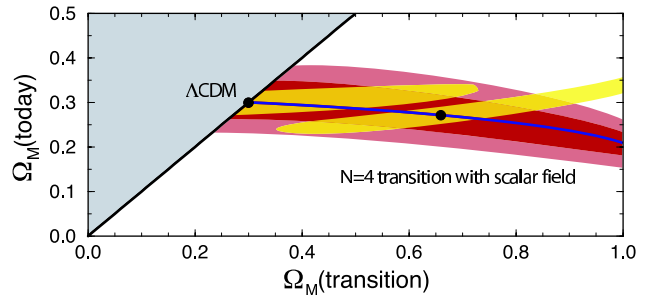}
\caption{Projected likelihood contours are shown for Planck CMB and SNAP SN
constraints on the $N=4$ gravitational transition model, including the
contribution of scalar field potential energy prior to the transition. Models
must lie below and to the right of the black line, outside the gray shaded
region, with $\Omega_m({\rm today}) \le \Omega_m({\rm transition})$. The dark
red contour shows the CMB 95$\%$ likelihood region, while the blue line
indicates the family of geometric degeneracy models. The lighter red contour
shows a weaker constraint from the CMB allowing for our uncertainty  in the
amount of spatial curvature. The black circles show the location of two specific
models: a $\Lambda$CDM model, and an $N=4$ gravitational transition. The Hubble
parameters for these two models have been chosen so that they have identical CMB
anisotropy spectra. The yellow contours show the SN 95$\%$ likelihood regions 
for these two underlying models. Wherease both models are viable at present, 
they are clearly distinguishable with future observations.} 
\label{figX}
\end{figure}

\noindent{\it Discussion.} 
The future prospects for distinguishing a gravitational transition from other
proposals for dark energy such as a cosmological constant are illustrated in
Fig.~\ref{figX}. Here, we consider the $N=4$ gravitational transition model
including the contribution of scalar field potential energy prior to the
transition. We assume a three-parameter family of models, consisting of
$\Omega_m$ today and at the transition, and the Hubble parameter $h$.  We have
calculated the $95\%$ confidence region based on the forecasts for CMB and SN
experiments, projected into the $\Omega_m({\rm transition}) - \Omega_m({\rm
today})$ plane. We use the proposed SNAP experiment
\cite{Kim:2003mq,unknown:2004ak} as the basis for forecasts to measure the
recent cosmological expansion history. The results are shown for two different
underlying models which are currently indistinguishable, $\Lambda$CDM with
$\Omega_m=0.3$ and the $N=4$ model with $\Omega_m=0.67,\,0.27$ at the transition
and today. In each case the Hubble parameter has been chosen so the two models
have identical CMB anisotropy, with the same matter density, $\Omega_m h^2$, and
angular-diameter distance to the last scattering surface, ${\cal D}_{CMB}$.
Next, we expect the Planck CMB \cite{Planck} experiment will determine ${\cal
D}_{CMB}$ to $0.2\%$ and $\Omega_m h^2$ to $0.9\%$, using temperature and
polarization data \cite{Hu:2001fb,Hu:2004kn}. A weaker constraint, due to our
lack of knowledge of the spatial curvature, whereby $\sigma(\ln \Omega_m
h^2)=0.018$ \cite{Eisenstein:1998hr} and $\sigma(\ln{\cal D}_{CMB})\approx
\frac{1}{4}\sigma(\ln \Omega_m h^2)$ \cite{Hu:2004kn}, is also shown. The figure
clearly indicates that these two sample cases are distinguishable. Other probes
of cosmic evolution, such as weak lensing and baryon acoustic oscillations can
further sharpen the distinction.

To summarize, we have introduced a new scenario which generalizes the vacuum
metamorpohsis model. In this scenario, gravitation undergoes a transition in
which an order parameter built out of curvature tensors freezes at a
constant value. After the transition, the matter content of the universe no
longer determines the cosmological evolution --- the expansion is ruled by the
value of the order parameter. The onset of cosmic acceleration is sudden, and
the effective dark energy equation-of-state is strongly negative, $w<-1$. 

We have demonstrated that a range of these models satisfy a number of the 
standard tests of cosmology. There is excellent motivation to study these
models further. The primary focus will be to analyze cosmological
perturbations and the impact on structure formation. We can expect to find
effects on the rate of growth of structure, the large-angle CMB anisotropy
pattern, and weak gravitational lensing. The other focus will be to examine
probes of the cosmic expansion suggesting the dark energy equation-of-state
sharply \cite{Alam:2003fg,Bassett:2002qu,Corasaniti:2002bw,Pogosian:2005ez}
dropped below $-1$ \cite{Kaplinghat:2003vf,Huterer:2004ch,Wang:2005ya}, a
distinct signature of this model.

\begin{acknowledgments} 

We thank Eric Linder for useful conversations. R.C. was supported in part by NSF
AST-0349213 at Dartmouth.   W.K. and L.P. were supported in part by the
Wisconsin Space Grant Consortium and by NSF PHY-0071044 at UWM. D.V. would like
to thank {\it Funda\c c\~ao de Amparo \`a Pesquisa do Estado de S\~ao Paulo}
(FAPESP) for full support.  

\end{acknowledgments}


\end{document}